\documentclass[sigconf]{acmart}

\usepackage{url}
\usepackage{breakurl}
\usepackage{balance}

\copyrightyear{2018}
\acmYear{2018} 
\setcopyright{iw3c2w3}
\acmConference[WWW '18 Companion]{The 2018 Web Conference Companion}{April 23--27, 2018}{Lyon, France}
\acmBooktitle{WWW '18 Companion: The 2018 Web Conference Companion, April 23--27, 2018, Lyon, France}
\acmPrice{}
\acmDOI{10.1145/3184558.3191612}
\acmISBN{978-1-4503-5640-4/18/04}

\begin{document}
\title{Gnirut: \\ The Trouble With Being Born Human In An Autonomous World}
\subtitle{}

\author{Luca Vigan\`{o}}
\affiliation{%
  \institution{King's College London}
  \streetaddress{Bush House, 30 Aldwych}
  \city{London}
  \state{UK}
  \postcode{WC2B 4BG}
}
\email{luca.vigano@kcl.ac.uk}

\author{Diego Sempreboni}
\affiliation{%
  \institution{King's College London}
  \streetaddress{Bush House, 30 Aldwych}
  \city{London}
  \state{UK}
  \postcode{WC2B 4BG}
}
\email{diego.sempreboni@kcl.ac.uk}

\renewcommand{\shortauthors}{L. Vigan\`o and D. Sempreboni}

\begin{abstract}
What if we delegated so much to autonomous AI and intelligent machines that They passed a law that forbids humans to carry out a number of professions? We conceive the plot of a new episode of Black Mirror 
to reflect on what might await us and how we can deal with such a future.
\end{abstract}

%
%



\maketitle

\section{Prologue: ``What do you want to be when you grow up?''}
EXT. MAIN STREET --- DAY. \\
An eight-year old child (C), one of the child's parents (P) and a woman they just met (W).
\begin{description}
\item[W:] And what do you want to be when you grow up?
\item[C:] Well, I...
\item[W:] A philosopher like your father?
\item[C:] I... I...
\item[P:] Come on, don't be shy. Answer the nice lady.
\item[W:] Or perhaps a painter like your mother?
\item[C:] I...
\item[P:] Come on, speak up! Please, forgive him. He's so shy. Come on!
\item[C:] I...
\item[P:] This is not ok. Not ok, you understand? The nice lady asked you a question and you should answer.
\item[W:] No worries, no worries.
\item[P:] Wait till we get home...
\item[W:] You'll tell me another time, ok?
\item[C:] ... a pilot! I want be a pilot!
\item[W:] Oh, but that's impossible, dear.
\item[C:] Why?
\item[W:] Because you can't.
\item[C:] I can't? Why?
\item[W:] It's not just you. We. We can't.
\item[C:] Why?
\item[P:] That's enough. Don't be a nuisance.
\item[W:] Don't be silly, no nuisance at all.
\item[C:] Why can't I be a pilot?
\item[P:] That's enough, I said!
\item[W:] Oh, no, no, please allow me to explain. There is a law.
\item[C:] A law?
\item[W:] A law that says we are not allowed to be pilots.
\item[C:] Why?
\item[W:] Because people were afraid.
\item[P:] People were afraid of flying, so They made a law.
\item[C:] How come people were afraid of flying? If at all, people should be afraid of falling.
\item[W:] Ha! That's clever.
\item[P:] People were too afraid that their time had come, you understand?
\item[C:] But if it had come, why worry? It had come anyway. On a plane or on the ground.
\item[W:] A-ha, and what if it's the pilot's time to go?
\item[P:] What if the pilot got sick, or decided to commit suicide?
\item[W:] What if the pilot made a mistake?
\item[C:] I wouldn't make any mistakes.
\item[W:] But of course you would. Of course. We all do.
\item[C:] I wouldn't get sick or kill myself. They could test me.
\item[W:] Oh, They did. They did test us. That's why They passed the law.
\item[P:] That's why They passed all those laws.
\item[W:] For our good.
\item[P:] Only for our good. That's why we have to obey.
\item[C:] But...
\item[P:] No but! You must learn to know your place.
\item[C:] But I want to be a pilot.
\item[W:] But you can't dear.
\item[P:] That's why They made a law. To protect us.
\item[W:] Pilots, doctors, surgeons, judges, construction builders, ... that's not for us anymore.
\item[P:] To protect us.
\item[C:] Yes, but why? Why can't I be a pilot?
\end{description}
We wrote this scene as the set-up of a future Black Mirror episode, in which ``They'' have passed laws that forbid humans to carry out a number of professions. How could this have happened?

\section{From the Human-to-AI Ratio ...}
The current recommendations (and, in some cases, regulations) require a \emph{many-to-1 human-to-robot ratio} (typically between \emph{2-to-1} and \emph{5-to-1}) for tasks that are particularly security-sensitive and/or safety-sensitive as they involve the protection of diverse critical assets such as data, systems, infrastructures and, ultimately, human life. Examples of such tasks are those carried out in military operations, such as 
drone strikes or rescue, demining or bomb-disposal operations --- studies carried out in the early 2000's in the context of military operations indicated the then current state of practice for aerial systems to be many-to-1 (as witnessed by the Global Hawk and Predator crews) and that ground robots are more effective with a 2-to-1 ratio~\cite{MurphyBurke10,BurkeMRLS04,BurkeMCR04}. 
Other, more ``civilian'', examples are tasks to be carried out in critical infrastructures or in specific automation environments and situations where human life or expensive assets might be in danger, such as remote surgeries or 
tasks carried out in power plants, in autonomous and assisted transportation systems, underwater~\cite{PalomerasCHKBCM16}, in space~\cite{Niles15}, or in harsh conditions, say due to excessive cold, heat or radiation~\cite{Fackler17}.

As stated by Murphy and Burke in 2010, the human-robot-inter\-act\-ion literature has (often) ignored safety, making the assumption that low human-to-robot ratios are desirable per se, and
\begin{quote}\em
often pursuing an arbitrary goal of 1-to-many based on expected advances in vehicle autonomy. \emph{\cite{MurphyBurke10}}
\end{quote}
Nowadays, this 1-to-many goal is not so arbitrary anymore: thanks to advances in robot autonomy, we appear to be not too far removed from a ratio of 1 human to 5 robots (\emph{1-to-5}) for all kinds of tasks, while ensuring safety and security (and privacy) of team, bystanders, and robots, as well as logical efficiency at the same time.

But why stop here? 
Solutions are currently being sought for full automation of systems by removing the human and transforming the operating space into a highly instrumented and controlled area for robots, as has been common in car manufacturing or even in places like Amazon warehouses (with carefully segregated spaces). The present reality is that in many dynamic and challenging environments we still cannot remove the human as there will be times an automated system is operating at the boundaries of its competence and might require human intervention. However, it is not utopian to imagine a near future reality in which the human-to-robot ratio 
has been reduced to \emph{0-to-5}, fulfilling the vision of a fully autonomous world in which robots carry out unmanned tasks. Actually, many tasks don't necessarily need a physical entity such as an arm or wheels to be carried out, but could be accomplished directly by some form of artificial intelligence, so we can speak more generally 
of a 0-to-5 \emph{human-to-AI ratio}.

But why stop here? 
Why not invert the ratio, and consider the case in which 1 human is supervised by one or more AIs? 
In 2017, Rachel Botsman carried out a small but enlightening experiment that involved her 3-year old daughter Grace and Amazon's Alexa, which she summarized in a New York Times Sunday Review Opinion
by observing that in the last few years, we have been (often passive) witnesses to 
\begin{quote}\em
a profound shift in our relationship with technology. For generations, our trust in it has gone no further than feeling confident the machine or mechanism will do what it's supposed or expected to do, nothing more, nothing less. We trust a washing machine to clean our clothes or an A.T.M. to dispense money, but we don't expect to form a relationship with them or call them by name.~\emph{\cite{Botsman17}}
\end{quote}
Now we can call them by name: Amazon's Alexa, Apple's Siri, Google's Google Assistant and Microsoft's Cortana are all intelligent personal assistants that recognize natural voice (in many different languages) without the requirement for keyboard input and support a wide range of user commands, ranging from answering questions to providing real-time information (such as news, weather forecast or traffic information), from making phone calls to compiling to-do lists, from setting alarms to playing music or audiobooks or streaming podcasts. Alexa can also control several smart devices using itself as a home automation system. It's the two-faced, sirenlike, beauty of the \emph{Internet of Things}, in which we silently enter an agreement to hand over the keys to our houses, to our appliances, to our data, to large parts of our lives:
\begin{quote}\em
Today, we're no longer trusting machines just to do something, but to decide what to do and when to do it. The next generation will grow up in an age where it's normal to be surrounded by autonomous agents, with or without cute names. The Alexas of the world will make a raft of decisions for my kids and others like them as they proceed through life --- everything from whether to have mac and cheese or a green bowl for dinner to the perfect gift for a friend's birthday to what to do to improve their mood or energy and even advice on whom they should date. In time, the question for them won't be, ``Should we trust robots?'' but ``Do we trust them too much?''~\emph{\cite{Botsman17}}
\end{quote}

\section{... to the AI-to-Human Ratio}
Our thesis in this paper is that we will not stop here. We will silently but willingly go for the full inversion of the ratio, from human-to-AI to AI-to-human, and hand over the vast majority of our choices and decisions, of our data and personal information, of our privacy, the vast majority of the different facets and aspects of our lives.
We won't stop at the 1-to-1 AI-to-human ratio exemplified by little Grace's naive, and somewhat cute, trust in Alexa:
\begin{quote}\em
With some trepidation, I watched my daughter gaily hand her decisions over. ``Alexa, what should I do today?'' Grace asked in her singsong voice on Day 3. It wasn't long before she was trusting her with the big choices. ``Alexa, what should I wear today? My pink or my sparkly dress?''~\emph{\cite{Botsman17}}
\end{quote}
We will go ``all in''.
The AI-to-human ratio will soon be \emph{1-to-some} (for instance, what if all of Botsman's family members, including the adults, had developed such a close relationship with Alexa?) and, ultimately, \emph{some-to-many}. 

This will happen gradually, with the change differentials initially too small to be noticed, or without us really realizing that the sequence of differentials over a longer period of time have caused a momentous change. This is somewhat reminiscent of the legendary social experiment involving 5 monkeys, a ladder and a banana in which, after a number of cold showers whenever one of the monkeys tried to reach for the banana, and after all 5 original monkeys have been stepwise replaced with 5 new monkeys, what is left is a group of 5 monkeys that, without ever having received a cold shower, continue to beat up any monkey who attempts to climb the ladder. The change has occurred gradually, one monkey at a time, but in the end a new status quo has been reached whose justification has long been forgotten: if it was possible to ask the monkeys why they beat up on all those who attempted to climb the ladder, their most likely answer would be ``because that's the way it's always been done around here.''\footnote{We wrote ``legendary'' as, apparently, the experiment was never really carried out but was rather inspired in part by the experiments of G.R. Stephenson, found in ``Cultural acquisition of a specific learned response among rhesus monkeys'', as well as experiments with chimpanzees conducted by Wolfgang Kohler in the 1920s. Over the years, it was pieced together to form the urban legend as it now stands, see \url{http://www.wisdompills.com/2014/05/28/the-famous-social-experiment-5-monkeys-a-ladder/} as well as Jeff Bridges'/President Jackson Evans' excellent rendition in the movie \emph{The Contender}, written and directed by Rod Lurie~\cite{Contender}.}
Will we end up being the monkeys in our own experiment towards a fully autonomous world?

There are already plenty of articles, news feeds and books explaining the future of employment and how intelligent machines will replace humans in many jobs in the context of the ``industrial revolution 4.0''~\cite{Kaplan15,Wakefield15,Ford16,Leonhard16,Bowcott17,Walsh17,Susskind17,Tegmark17}.
It is not just that many boring or strenuous jobs (such as taxi driver~\cite{autodrive}, factory worker~\cite{Davis18}, etc.) are under threat of automation and will result in technological unemployment --- this is often referred to as the \emph{economic singularity}~\cite{Chace16} that will bring us one step closer to the \emph{technological singularity} in which ordinary humans will someday be overtaken by artificially intelligent machines or cognitively enhanced biological intelligence~\cite{Shanahan15}. In a laudable attempt to safeguard human safety, we will also likely soon legislate that some jobs 
should be carried out only by intelligent machines. This won't be limited to what is already happening (as we remarked above, robots have, at least partially, taken over dangerous operations such as demining, bomb-disposal, or nuclear meltdown inspections). With the advances in the automation of decision making, planning and autonomy, this will also encompass professions in which human nature and subjectivity might slow down reaction time or adversely affect the end-result. 

Think, for example, about the story of the ``Miracle on the Hudson'': the US Airways Flight 1549, in the climbout after takeoff from New York City's LaGuardia Airport on January 15, 2009, struck a flock of Canada geese and consequently lost all engine power. Unable to reach any airport, the pilot Chesley ``Sully'' Sullenberger and his co-pilot Jeffrey Skiles glided the plane to a ditching in the Hudson River off Midtown Manhattan. All 155 people aboard were rescued by nearby boats and there were only few serious injuries. As masterfully portrayed in the movie \emph{Sully} directed by Clint Eastwood from a screenplay by Todd Komarnicki (based on the book \emph{Highest Duty} by Sullenberger and Jeffrey Zaslow), Sullenberger and Skiles were subject to an investigation by the National Transportation Safety Board, which initially seemed to conclude that they could have safely landed in one of the nearby airports instead of attempting a one-in-a-million landing in the river. A number of pilots were asked to carry out human-piloted simulations of the accident which all showed that it was actually possible to land safely in one of the nearby airports. This is Tom Hank's/Sully's reply
\begin{quote}\em
...you're still not taking into account the human factor. These pilots are not reacting like human beings. Like people who are experiencing this for the first time. 
... \\
You have allowed no time for analysis and decision making. And with these sims, you have taken all the humanity out of the cockpit.~\emph{\cite{Sully}}
\end{quote}
Reaction-decision time is then set at thirty-five seconds and the simulations are run again... and all result in a crash. With that delay, attempting to land in the Hudson was indeed the only option. An intelligent machine, on the other hand, might have required a much lower 
reaction-decision time and safely made it to a runway.

To overcome, and prevent, similar problems, the rise in AI will alter also legal frameworks. As observed by Bowcott in his analysis of a report by the International Bar Association~\cite{Bowcott17},
\begin{quote}\em
Among the professions deemed most likely to disappear are accountants, court clerks and `desk officers at fiscal authorities'. ... 
Even some lawyers risk becoming unemployed. ``An intelligent algorithm went through the European Court of Human Rights' decisions and found patterns in the text,'' the report records. ``Having learned from these cases, the algorithm was able to predict the outcome of other cases with 79\% accuracy ... According to a study conducted by [the auditing firm] Deloitte, 100,000 jobs in the English legal sector will be automated in the next 20 years.''
\end{quote}

But why stop here?
Why not hand over the actual legislative power to intelligent machines, which, after all, have the ability to take better and more consistent decisions? ``They'' might then legislate that the list of professions that should not be accessible to humans ought be extended to include also all kinds of drivers and pilots (no more Sully, then), doctors, surgeons, accountants, lawyers, judges, lawmakers, soldiers, ... you get the drift.

What will be left for us humans to do?
Well, first of all, we will hope that the system doesn't go \emph{Skynet} on us\footnote{\emph{Skynet} is the fictional AI that serves as the main antagonist in the movies of the \emph{Terminator} franchise~\cite{Terminator}. 
Skynet came to the logical conclusion that all of humanity would attempt to destroy it. In order to continue fulfilling its programming mandates of ``safeguarding the world'' and to defend itself against humanity, Skynet launched a series of nuclear attacks killing over three billion people and gathered a slave labor force from surviving humans.}, but then we will also be faced with the problem of what to do with our lives. Will we end up fat and lazy, although potentially happy, like the humans in \emph{Wall$\cdot{}$E}~\cite{WallE}? Will we all be rich and bored, and have lives of leisure while the machines are taking the sweat~\cite{Walsh17}?
Will we succumb to pessimism like Cioran, whose best-known work~\cite{Cioran} inspired the subtitle of this paper?
\begin{quote}
\begin{description}\em
\item[---] What do you do from morning to night?
\item[---] I endure myself.
\end{description}
\end{quote}
So, what will be left for us to do? Actually, Cioran himself (unknowingly) suggests an answer:
\begin{quote}\em
A zoologist who observed gorillas in their native habitat was amazed by the uniformity of their life and their vast idleness. Hours and hours without doing anything. Was boredom unknown to them? This is indeed a question raised by a human, a busy ape. ... 
Man alone, in nature, is incapable of enduring monotony, man alone wants something to happen at all costs --- something, anything...~\emph{\cite{Cioran}}
\end{quote}
A common distinguishing trait of us human beings with respect to animals, monkeys, apes and primates, and possibly with respect to present and future AI, is our desire for more, our insatiable curiosity.\footnote{As Albert Einstein 
said, ``I have no special talents, I am only passionately curious.''} We desire what we don't have, and in some cases what we can't have, and we strategize and make long-term plans to achieve what we desire. In the scene at the beginning of this paper, the child C wants to become a pilot. C doesn't care that the laws forbid it, C wants to fly. C asks ``why?'', C is the monkey that challenges the status quo, the way things are done around here. 

What can C do then? If it really wishes to become a pilot, C will muster up a plan. C will be prepared to lie to achieve what it wants. Perhaps inspired by Yentl, the Jewish girl who disguises herself as a boy to enter religious training in Isaac Bashevis Singer's short story and play \emph{Yentl, the Yeshiva Boy}~\cite{Yentl} as well as in Barbra Streisand's eponymous movie~\cite{Yentlmovie} (but see also similar plotlines in \emph{Tootsie}~\cite{Tootsie}, \emph{Albert Nobbs}~\cite{Nobbs} and many more movies, books and plays going back to Greek mythology and actually even earlier than that). 

\section{Can life imitate AI?}

C could attempt to disguise itself as an AI to enter the ``pilot academy'' of this futuristic, but not utopian, fully autonomous world. Will C be able to pass the (to-be-developed) \emph{Gnirut Test}, a fully reverse Turing test that will involve an AI judge and a human subject  which attempts to appear artificial? The underlying presumption here is that an AI subject will always be judged artificial, and a human is said to ``pass the Gnirut test'' if it is also judged artificial. 

The idea of a deceptive machine is fundamental to AI research and has been present in the AI literature since Turing introduced the \emph{Imitation Game}, so why not consider a deceptive human in our plot?
Will C be able to deceive the AI judge? Difficult, probably impossible, but we wish to believe that at least in our Black Mirror episode C will stand a chance.\footnote{Or maybe the AI judge will fall in love with C and help it to go unnoticed like Rick Deckard does with Rachel in \emph{Blade Runner}~\cite{BladeRunner} (although the original novella by Philip K. Dick~\cite{Dick} actually followed a slightly different storyline).} 

Advances in \emph{Deception Theory} might provide C with a range of techniques to fool the judge, especially when coupled with advances in the \emph{Theory of Mind}, which investigates the creation of intelligent machines that are able to model other agents' minds. Maybe then C will be able not only to pass as an AI but also to lower its reaction-decision time to that of an intelligent machine in case of an accident to the plane it is piloting. C will hopefully be able to leverage on the (to-be-developed) \emph{Theory of Artificial Mind} to plot its disguise in what could be an entertaining and, as usual, thought-provoking new episode of Black Mirror aptly titled \emph{Gnirut}.

\section{Epilogue}
Like most episodes of Black Mirror, our own Gnirut would also be discomforting and scary. It would ask a lot of ``What ifs?'' and act as a cautionary tale, putting us in front of that black mirror that we can find 
\begin{quote}\em
on every wall, on every desk, in the palm of every hand: the cold, shiny screen of a TV, a monitor, a smartphone.~\emph{\cite{Brooker}}
\end{quote}
Should we be scared by the reflection that we see? Of course, we should, at least a little. Should we retire Alexa to the closet like Botsman tells us that she did after the experiment with her daughter~\cite{Botsman17}? Our answer is no, we should not. 

Rather than halting technological progress, we should make sure to accompany it, bringing together, in a multi-disciplinary supervision effort, teams of experts including AI experts, informaticians, mathematical and social scientists, psychologists, lawmakers and more. 
We should be careful who we trust~\cite{BotsmanTrust17} and, above all, we should not lose our faith in expertise~\cite{Nichols17}. 
Only then will the ``magnificent and progressive fate of the human race''~\cite{Leopardi} stand a chance to materialize in its full glory.

\begin{acks}
  We thank Daniele Magazzeni for many interesting discussions (and for allowing us to paraphrase a couple of his still unpublished sentences).
\end{acks}

\balance


\end{document}